\documentstyle[pra,aps,floats]{revtex}

\def\C{{\rm\kern.24em
    \vrule width.02em height1.4ex depth-.05ex
    \kern-.26em C}}
\def\F{{\rm I\kern-.25em F}}
\def\Z{{\rm\kern.26em
    \vrule width.02em height0.5ex depth 0ex
    \kern.04em
    \vrule width.02em height1.47ex depth-1ex
    \kern-.34em Z}}

\def\bra#1{\left<#1\right|}
\def\ket#1{\left|#1\right>}

\def\qed{\rightline{$\Box$}}

\def\trace{{\rm tr}}

\title{Computing Local Invariants of Qubit Systems}

\author{Markus Grassl\thanks{e-mail: {\protect\tt
grassl@ira.uka.de}}, Martin R\"otteler\thanks{e-mail: {\protect\tt
roettele@ira.uka.de}}, and Thomas Beth\thanks{e-mail: {\protect\tt
EISS\_Office@ira.uka.de}}}
\address{Institut f{\"u}r Algorithmen und Kognitive Systeme, Universit{\"a}t
Karlsruhe, Am Fasanengarten 5, D-76\,128 Karlsruhe, Germany.}
\date{May 29, 1998}

\begin{document}

\twocolumn
\narrowtext
\maketitle
\begin{abstract}%
We investigate means to describe the non-local properties of quantum
systems and to test if two quantum systems are locally equivalent. For
this we consider quantum systems that consist of several subsystems,
especially multiple qubits, i.\,e., systems consisting of subsystems
of dimension two. We compute invariant polynomials, i.\,e., polynomial
functions of the entries of the density operator which are invariant
under local unitary operations.

As an example, we consider a system of two qubits. We compute the
Molien series for the corresponding representation which gives
information about the number of linearly independent
invariants. Furthermore, we present a set of polynomials which
generate all invariants (at least) up to degree 23. Finally, the use
of invariants to check whether two density operators are locally
equivalent is demonstrated.
\end{abstract}

\section{Introduction}
Non-locality is one of the astonishing phenomena in quantum
mechanics. Well-known examples are EPR pairs \cite{EPR35} and the GHZ
state \cite{GHZ89}. States of quantum codes contradict local realism,
too \cite{DiVincenzoPeres97}. One common feature of these states is
that the non-local properties do not change under local
transformations, i.\,e., unitary operations acting independently on
each of the subsystems. Thus, any function invariant under local
unitary transformations (LUT) can be used to describe these non-local
properties \cite{SchMah95,SchMah96}. Here we study polynomials that
are invariant under local unitary transformations. Among these, there
are e.\,g. the coefficients of the characteristic polynomial of a
density operator or of the reduced density operators. The paper
extends the work of Rains~\cite{Rains97} in which he showed how, in
principle, all polynomial invariants can be computed. We present
further reductions of complexity which make the computation of
polynomial invariants feasible at least for small systems.

The paper is organized as follows: In Section~\ref{PolyInv} we
consider the linear action and the action by conjugation of matrix
groups on polynomials. Then we establish a connection between
invariant polynomials and the algebra of matrices commuting with all
elements of the group. A physical interpretation of the invariant
polynomials is given by relating them to some observables. Mainly
classical results for these algebras are recalled in
Section~\ref{MatAlgebra}. In Section~\ref{Trees} a method to construct
a vector space basis of these algebras is presented for the case of
two-dimensional subsystems. Furthermore, we present results that imply
a further reduction of the complexity to compute all invariants. The
special situation of pure states and subspaces is considered in
Section~\ref{PureSubspace}. In Section~\ref{TwoQubit} we compute the
Molien series and a set of invariants for a two qubit system. We
conclude in Section~\ref{Examples} with examples for the application
of these invariants.

\section{Polynomial invariants}\label{PolyInv}
\subsection{Operation on polynomials}
The group $GL(n,\F)$ of invertible $n\times n$ matrices over the
field $\F$ operates on the polynomials
$p(x_1,\ldots,x_n)\in\F[x_1,\ldots,x_n]$ in the following manner:
\begin{eqnarray}
&&p^g(x_1,\ldots,x_n):=p(\widetilde{x}_1,\ldots,\widetilde{x}_n)
\nonumber\\
&&\qquad\mbox{where $(\widetilde{x}_1,\ldots,\widetilde{x}_n)
                    :=(x_1,\ldots,x_n)^ g$}\label{lin_operation}
\end{eqnarray}
i.\,e., each variable is substituted by the linear combination
obtained by multiplying the vector of variables by the group element
$g\in GL(n,\F)$.

On $n\times n$ matrices the group $GL(n,\F)$ acts by conjugation.
Hence polynomials $f(\rho_{ij})=f(\rho_{11},\ldots,\rho_{nn})$ in the
entries $\rho_{ij}$ of a $n\times n$ density operator $\rho$, are then
acted upon by conjugation, i.\,e.,
\begin{equation}\label{conj_operation}
f^g(\rho_{ij})=f(\widetilde{\rho}_{ij})
\qquad\mbox{where $\widetilde{\rho}:=\rho^g=g^{-1}\cdot\rho\cdot g.$}
\end{equation}

Given a subgroup $G\le GL(n,\F)$, we are interested in polynomials
that are fixed by all elements of $G$ under the action defined by
either (\ref{lin_operation}) or (\ref{conj_operation}). These
invariant polynomials (just called {\em invariants}) form an algebra
over the field $\F$ since any linear combination and any product of
invariants is invariant under the action of the group, too. It is
sufficient to study homogeneous polynomials as each homogeneous
polynomial of degree $k$ remains homogeneous of the same degree under
the operation of $G$ and every polynomial can be decomposed additively
into its homogeneous components.

For the class of so-called {\em reductive} groups (e.\,g.{} finite
groups, unitary groups) the invariant ring is finitely generated
(cf.~\cite{Springer}), i.\,e., every invariant can be expressed in
terms of some algebra generators. These so-called {\em fundamental
invariants} can be chosen to be homogeneous polynomials of small
degree. Under this assertion the task is to find a system of
fundamental invariants such that any other invariant can be expressed
as a polynomial of these. In what follows we focus on this task for
invariants under the action of tensor products of unitary groups on
density operators by conjugation given by (\ref{conj_operation}).
\subsection{Invariant matrices}
Instead of studying the invariant polynomials directly, we use the
relation between homogeneous polynomials in the entries of a density
operator $\rho$ and constant matrices.

{\em Lemma \ref{poly_mat}.}
For every homogeneous polynomial $f$ of degree $k$ in the entries of
the density operator $\rho$ there exists a matrix $F$ such that
\begin{equation}\label{trace_poly}
f(\rho_{ij})=f_F(\rho_{ij})=\trace\left(F\cdot \rho^{\otimes k}\right).
\end{equation}

{\em Proof.} This follows directly from the fact that the matrix
$\rho^{\otimes k}$ contains all monomials in the variables $\rho_{ij}$
of degree $k$.

\qed
Next, we characterize invariant polynomials in terms of the
corresponding matrices.

{\em Theorem \ref{rel_poly_mat}.}
A homogeneous polynomial $f$ of degree $k$ in the entries of the
density operator $\rho$ is invariant under the operation of a compact
group $G\le GL(n,\F)$ by conjugation if and only if $f=f_F$ for a
matrix $F$ that is invariant under conjugation by
$(g^{-1})^{\otimes k}$ (equivalently, iff the matrix $F$ commutes
with $g^{\otimes k}$) for all $g\in G$.

{\em Proof.}
Conjugation of $\rho$ by $g$ corresponds to conjugation of $F$ by
$(g^{-1})^{\otimes k}$ as shown by the following calculation:
\begin{eqnarray*}
f_F\left((g^{-1}\cdot\rho\cdot g)_{ij}\right)
&=&\trace\left(F\cdot(g^{-1}\cdot\rho\cdot g)^{\otimes k}\right)\\
&=&\trace\left(F\cdot(g^{-1})^{\otimes k}
            \cdot\rho^{\otimes k}
            \cdot g^{\otimes k}\right)\\
&=&\trace(
 \underbrace{g^{\otimes k}\cdot F\cdot(g^{-1})^{\otimes k}}_{\widetilde{F}}
   \cdot \rho^{\otimes k})\\
&=&f_{\widetilde{F}}(\rho_{ij}).
\end{eqnarray*}
If $F$ commutes with $g^{\otimes k}$, then the equality of
$\widetilde{F}$ and $F$ implies
$f_F\left((g^{-1}\rho g)_{ij}\right)=f_{F}(\rho_{ij})$.

If on the other hand $f$ is invariant under the operation of $G$, for
any matrix $F$ with $f=f_F$ we have $f_F\left((g^{-1} \rho
g)_{ij}\right)=f_{F}(\rho_{ij})$. Since the group $G$ is compact, we
can average over the group (cf.~\cite{Weyl39}) and obtain the matrix
\begin{equation}\label{int_matrix}
\overline{F}=
  \int_{g\in G} 
     \left(g^{\otimes k}\cdot F\cdot (g^{-1})^{\otimes k}\right)
   \;d\mu_G(g).
\end{equation}
By construction, $\overline{F}$ is invariant under conjugation by
$(g^{-1})^{\otimes k}$ and furthermore $f=f_{\overline{F}}$.

\qed

Using this theorem and Lemma~\ref{poly_mat}, we are in principle able
to compute invariants of the group $G$ starting from any matrix $F$
and computing a matrix $\overline{F}$ that commutes with $g^{\otimes
k}$ for all $g\in G$. But in practice, the integration
(\ref{int_matrix}) is very difficult to perform. In
Section~\ref{MatAlgebra} we will present a method to calculate the
matrices $F$ directly without integration.

\subsection{Physical interpretation of the polynomial
invariants}\label{Physical}
Although we do not have full insight in the physical interpretation of
the polynomial invariants yet, we will relate them to some observables.

Recall from Eq.~(\ref{trace_poly}) that all polynomial invariants of
degree $k$ can be written as $f(\rho_{ij})_F=\trace\left(F\cdot
\rho^{\otimes k}\right)$. From $F$ we construct two Hermitian
operators
$$
M_1:=F+F^\dagger\quad\mbox{and}\quad M_2:=i\,F-i\,F^\dagger.
$$
Both $M_1$ and $M_2$ commute with $g^{\otimes k}$ for all
$g=U_1\otimes\ldots\otimes U_N\in U(n)^{\otimes N}$ since $F$ (and
thus $F^\dagger$) commutes with $g^{\otimes k}$
(cf.~Theorem~\ref{rel_poly_mat}). Hence the (real) mean values
$$
\left<M_1\right>:=\trace\left(M_1\cdot\rho^{\otimes k}\right)
\quad\mbox{and}\quad
\left<M_2\right>:=\trace\left(M_2\cdot\rho^{\otimes k}\right)
$$
are also invariant under local unitary transformation. In principle
they can be obtained by joint measurements of $k$ copies of the
quantum system with density operator $\rho$.

\section{Invariant algebras}\label{MatAlgebra}

In order to compute all homogeneous invariants of degree $k$ it is
sufficient to know the algebra of matrices that commute with
$g^{\otimes k}$ for all $g\in G$. Such algebras have been studied
e.\,g.{} by Brauer for many classes of groups
(cf.~\cite{Brauer37}). For the unitary group and tensor products of
unitary groups, we have the following theorems and corollaries:

{\em Theorem \ref{Brauer}. (Brauer)}
The matrix algebra ${\cal A}_{n,k}$ of matrices that commute with
any matrix $U^{\otimes k}$ for $U\in U(n)$ is generated by the
representation $T_{n,k}\colon S_k\rightarrow GL(n^k,\C)$ of the
symmetric group $S_k$ that operates on the tensor product space
$(\C^n)^{\otimes k}=V_1\otimes\ldots\otimes V_k$ by permuting the $k$
spaces $V_i$ of dimension $n$.

This result extends to tensor products of unitary groups:

{\em Corollary \ref{coro4}.}
The algebra of matrices that commute with any matrix $U_1^{\otimes
k}\otimes\ldots\otimes U_N^{\otimes k}$ for $U_i\in U(n_i)$ is given
by the tensor product of the algebras ${\cal A}_{n_i,k}$.

To obtain the ``natural'' ordering of the tensor factors, we have to
conjugate the matrices by a permutation matrix:

{\em Corollary \ref{multiAlgebra}.}
The algebra ${\cal A}_{n,k}^{(N)}$ of matrices that commute with any
matrix $\left(U_1\otimes\ldots\otimes U_N\right)^{\otimes k}$ for
$U_i\in U(n)$ is conjugated to the $N$-fold tensor product of the
algebra ${\cal A}_{n,k}$, i.\,e.,
\begin{equation}\label{TensorAlgebra}
{\cal A}_{n,k}^{(N)}=\sigma
                 \left({\cal A}_{n,k}\right)^{\otimes N}
                 \sigma^{-1},
\qquad\mbox{$\sigma:=T_{n,kN}(\tau)$}.
\end{equation}
Here $\tau$ is the permutation that exchanges the macro- and
micro-coordinates according to the isomorphism between the tensor
product spaces
$$
\left(V^{\otimes N}\right)^{\otimes k}
\quad\mbox{and}\quad
\left(V^{\otimes k}\right)^{\otimes N}
\qquad\mbox{(where $V=\C^n$)}.
$$
As a permutation on $\{1,\ldots,k\cdot N\}$, $\tau$ maps $ak+b+1$
to $bN+a+1$ (for $a=0,\ldots,N-1$, $b=0,\ldots,k-1$).

(The reader familiar with the theory of fast Fourier transformations
will recognize the similarity to the ``bit reversal permutation''
\cite{FFT}.)

For the special situation of qubit systems, i.\,e., the group $U(2)$,
the dimension of the algebra is given by the following theorem:

{\em Theorem \ref{dimRep}.}
The vector space dimension of the algebra ${\cal A}_{2,k}$ is given
by the Catalan number~\cite{ConcreteMath}
$$
C(k)=\frac{1}{k+1}{{2k} \choose k}.
$$

{\em Proof.} This result is derived in \cite{Roetteler97} from a
theorem of Weyl~\cite{Weyl39}.\hfill$\Box$

Note that the algebra ${\cal A}_{2,k}$ was defined by the $k!$
matrices $T_{2,k}(\pi)$ for $\pi\in S_k$. As an algebra, ${\cal
A}_{2,k}$ is generated by the image of the generators of $S_k$,
i.\,e., by only two matrices. Theorem~\ref{dimRep} states that, even
as a vector space, less than $k!$ matrices are sufficient.

\section{Binary trees, permutations, and algebras}\label{Trees}
\subsection{One qubit}
The mapping (\ref{trace_poly}) from invariant matrices to invariant
homogeneous polynomials is a vector space homomorphism. Thus, in order
to compute all linearly independent homogeneous invariants of degree
$k$ for tensor products of the group $U(2)$, it is sufficient to
consider a vector space basis of the algebra ${\cal A}_{2,k}$. Such a
basis can be constructed starting from binary trees with $k$ nodes,
mapping them to permutations of $k$ letters, and finally obtaining
matrices via the representation $T_{2,k}$. The construction resembles
some of the many beautiful combinatorial properties of Catalan numbers
(cf.~\cite{ConcreteMath,Lueneburg}).

\begin{figure}[hbt]
\centerline{\unitlength1.5pt
\begin{picture}(25,20)(-5,0)
\put(20,20){\vector(-1,-1){9}}
\put(20,20){\makebox(0,0){$\bullet$}}
\put(18,20){\makebox(0,0)[r]{\scriptsize 1}}
\put(10,10){\vector(-1,-1){9}}
\put(10,10){\makebox(0,0){$\bullet$}}
\put(8,10){\makebox(0,0)[r]{\scriptsize 2}}
\put(0,0){\makebox(0,0){$\bullet$}}
\put(-2,0){\makebox(0,0)[r]{\scriptsize 3}}
\end{picture}
\begin{picture}(20,20)(5,0)
\put(20,20){\vector(-1,-1){9}}
\put(20,20){\makebox(0,0){$\bullet$}}
\put(18,20){\makebox(0,0)[r]{\scriptsize 1}}
\put(10,10){\vector(1,-1){9}}
\put(10,10){\makebox(0,0){$\bullet$}}
\put(8,10){\makebox(0,0)[r]{\scriptsize 2}}
\put(20,0){\makebox(0,0){$\bullet$}}
\put(22,0){\makebox(0,0)[l]{\scriptsize 3}}
\end{picture}
\begin{picture}(30,20)(-5,0)
\put(10,20){\vector(-1,-1){9}}
\put(10,20){\vector(1,-1){9}}
\put(10,20){\makebox(0,0){$\bullet$}}
\put(8,20){\makebox(0,0)[r]{\scriptsize 1}}
\put(0,10){\makebox(0,0){$\bullet$}}
\put(-2,10){\makebox(0,0)[r]{\scriptsize 2}}
\put(20,10){\makebox(0,0){$\bullet$}}
\put(22,10){\makebox(0,0)[l]{\scriptsize 3}}
\end{picture}
\begin{picture}(25,20)
\put(10,20){\vector(1,-1){9}}
\put(10,20){\makebox(0,0){$\bullet$}}
\put(8,20){\makebox(0,0)[r]{\scriptsize 1}}
\put(20,10){\makebox(0,0){$\bullet$}}
\put(22,10){\makebox(0,0)[l]{\scriptsize 2}}
\put(20,10){\vector(-1,-1){9}}
\put(10,0){\makebox(0,0){$\bullet$}}
\put(8,0){\makebox(0,0)[r]{\scriptsize 3}}
\end{picture}
\begin{picture}(30,20)(-5,0)
\put(0,20){\vector(1,-1){9}}
\put(0,20){\makebox(0,0){$\bullet$}}
\put(-2,20){\makebox(0,0)[r]{\scriptsize 1}}
\put(10,10){\makebox(0,0){$\bullet$}}
\put(12,10){\makebox(0,0)[l]{\scriptsize 2}}
\put(10,10){\vector(1,-1){9}}
\put(20,0){\makebox(0,0){$\bullet$}}
\put(22,0){\makebox(0,0)[l]{\scriptsize 3}}
\end{picture}
}
\bigskip
\caption{The five distinct ordered binary trees with three nodes.
\label{BinTrees}} 
\end{figure}
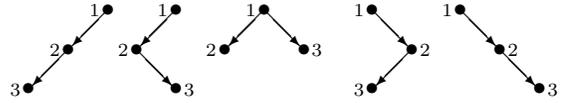

Let $B_k$ denote a labeled ordered binary tree with $k$ nodes, i.\,e.,
each node in the tree but the root has a father, and each node in the
tree has at most one left and at most one right son. The labeling of
the $k$ nodes of the tree with the numbers $\{1,\ldots,k\}$ is
obtained by traversing the nodes in the order root, left sub-tree,
right sub-tree. FIG.~\ref{BinTrees} shows all distinct binary trees
with three nodes labeled in that manner.

A {\em maximal right path} in the binary tree $B_k$ is a sequence of
nodes $(r_0,r_1,\ldots,r_j)$ such that each of the nodes $r_{i+1}$ is
the right son of the node $r_i$, $r_0$ is not the right son of any
node, and $r_j$ has no right son.

Given the set ${\cal R}(B_k)$ of all maximal right paths of a binary
tree $B_k$ we define a permutation $\pi(B_k)\in S_k$ by the product of
cycles
\begin{equation}\label{tree2perm}
\pi(B_k)=\prod_{(r_0,r_1,\ldots,r_{j_i})\in{\cal R}(B_k)}
  (r_0\;r_1\;\ldots\;r_{j_i}).
\end{equation}
E.\,g., for the trees of FIG.~\ref{BinTrees} we get the five
permutations $(1)(2)(3)$, $(1)(2\;3)$, $(1\;3)(2)$, $(1\;2)(3)$, and
$(1\;2\;3)$.

Let ${\cal B}_k$ denote the set of all distinct ordered binary trees
with $k$ nodes labeled in the manner described before, and be ${\cal
P}_k=\pi({\cal B}_k)$ the set of permutations obtained by the mapping
$\pi$. Using this notation, we can formulate the following theorem:

{\em Theorem \ref{CatalanBasis}.} 
The set of matrices 
${\cal M}_k:=T_{2,k}({\cal P}_k)=\{T_{2,k}(\pi(B)): B \in{\cal B}_k\}$
forms a vector space basis of the algebra ${\cal A}_{2,k}$.

{\em Proof.} 
For $k=0$ and $k=1$ the statement is obviously true.

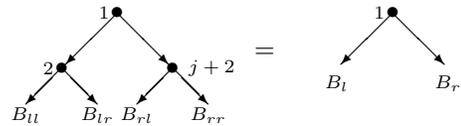
\begin{figure}[hbt]
\centerline{\unitlength1.5pt
\begin{picture}(50,40)(-5,0)
\put(20,40){\vector(-1,-1){13}}
\put(20,40){\vector(1,-1){13}}
\put(20,40){\makebox(0,0){$\bullet$}}
\put(18,40){\makebox(0,0)[r]{\scriptsize 1}}
\put(6,26){\makebox(0,0){$\bullet$}}
\put(4,26){\makebox(0,0)[r]{\scriptsize 2}}
\put(34,26){\makebox(0,0){$\bullet$}}
\put(38,26){\makebox(0,0)[l]{\scriptsize $j+2$}}
\put(6,26){\vector(-1,-1){9}}
\put(-3,16){\makebox(0,0)[t]{\scriptsize $B_{ll}$}}
\put(6,26){\vector(1,-1){9}}
\put(15,16){\makebox(0,0)[t]{\scriptsize $B_{lr}$}}
\put(34,26){\vector(-1,-1){9}}
\put(25,16){\makebox(0,0)[t]{\scriptsize $B_{rl}$}}
\put(34,26){\vector(1,-1){9}}
\put(43,16){\makebox(0,0)[t]{\scriptsize $B_{rr}$}}
\end{picture}
\begin{picture}(20,40)
\put(10,30){\makebox(0,0){$=$}}
\end{picture}
\begin{picture}(40,40)
\put(20,40){\vector(-1,-1){13}}
\put(20,40){\vector(1,-1){13}}
\put(20,40){\makebox(0,0){$\bullet$}}
\put(18,40){\makebox(0,0)[r]{\scriptsize 1}}
\put(6,24){\makebox(0,0)[t]{\scriptsize $B_l$}}
\put(34,24){\makebox(0,0)[t]{\scriptsize $B_r$}}
\end{picture}
}
\caption{General form of a tree in the class ${\cal B}_{k,j}$. The
sub-trees $B_{ll}$, $B_{lr}$, $B_{rl}$, and $B_{rr}$ might be
empty. Furthermore, for $j=0$ (or $j=k-1$) the left sub-tree $B_l$
(or the right sub-tree $B_r$, resp.) is empty. \label{genClassTree}}
\end{figure}

For $k>1$, we partition the set ${\cal B}_k$ of binary trees with $k$
nodes into $k$ classes ${\cal B}_{k,j}$ ($j=0,\ldots,k-1$). The class
${\cal B}_{k,j}$ consists of all trees with $j$ nodes in the left
sub-tree of the root. The general form of a tree in the class ${\cal
B}_{k,j}$ is shown in FIG.~\ref{genClassTree}.

For $j<k-1$ in each of the trees $B\in{\cal B}_{k,j}$, $j+2$ is a
right son of $1$, and thus $\pi(B)$ maps $1$ to $j+2 k$. For $j=k-1$,
the root $1$ has no right son and thus $\pi(B)$ fixes $1$. To combine
these two cases, we identify $k+1$ and $1$. Hence $T_{2,k}(\pi(B))$
maps $\ket{\bbox{e}_1}$ to $\ket{\bbox{e}_{j+2}}$ where
$\ket{\bbox{e}_i}=\ket{0}^{\otimes{i-1}}\ket{1}\ket{0}^{\otimes{k-i}}$.
This shows that for $B\in{\cal B}_{k,j}$
$$
\trace\bigl(\ket{\bbox{e}_1}\bra{\bbox{e}_{j'+2}} T_{2,k}(\pi(B))\bigr)
=\delta_{j,j'}.
$$
Therefore, for $j\ne j'$ the matrices in the sets $T_{2,k}(\pi({\cal
B}_{k,j}))$ and $T_{2,k}(\pi({\cal B}_{k,j'}))$ are mutually linearly
independent.

For fixed $j$, each permutation $\pi(B)$ for a tree $B\in{\cal
B}_{k,j}$ with left and right sub-trees $B_l$ and $B_r$ (see
FIG.~\ref{genClassTree}) can be written in the form
$$
\pi(B)=(1\;{j+2})\cdot\pi(B_l)\cdot\pi(B_r)
$$
where the permutations $\pi_l=\pi(B_l)$ and $\pi_r=\pi(B_r)$ operate
on the sets $\{2,\ldots,j+1\}$ and $\{j+2,\ldots,k\}$, resp. The
corresponding representations are ``shifted'' tensor products, i.\,e.,
\begin{eqnarray*}
T_{2,k}(\pi(B))&=&
  T_{2,k}\bigl((1\;{j+2})\bigr)\cdot
    \left(\openone_{2^{j+1}}\otimes T_{2,{k-j-1}}(\pi_r')\right)\\
&&\quad\cdot\left(\openone_{2^1}\otimes T_{2,j}(\pi_l')
             \otimes\openone_{2^{k-j-1}}\right)\\
&=&  T_{2,k}\bigl((1\;{j+2})\bigr)\\
&&\quad
\cdot\left(\openone_{2^1}\otimes T_{2,j}(\pi_l')
  \otimes T_{2,{k-j-1}}(\pi_r')\right)
\end{eqnarray*}
where $\pi_l'\in S_j$ and $\pi_r'\in S_{k-j-1}$ are obtained by
relabeling. By the induction hypothesis, the matrices corresponding to
the sub-trees $B_l$ and $B_r$ are linearly independent for different
trees. Thus we have shown that the matrices in ${\cal M}_k$ are
linearly independent. It remains to show that they form a basis of
${\cal A}_{2,k}$. But this follows from Theorem~\ref{dimRep} together
with the fact that there are exactly $C(k)$ different ordered binary
trees with $k$ nodes (cf.~\cite[p.~389]{Knuth73}).

\qed

\subsection{Multiple qubits}
In order to compute a basis of the algebra ${\cal A}_{n,k}^{(N)}$ for
an $N$ particle system, we define the following representation
$T^{(N)}_{n,k}\colon \left(S_k\right)^N\rightarrow GL(n^{kN},\C)$ of
the $N$-fold direct product of the symmetric group $S_k$:
$$
(\pi_1,\ldots,\pi_N)\mapsto
\sigma\cdot\left(T_{n,k}(\pi_1)\otimes\ldots\otimes
T_{n,k}(\pi_N)\right)\cdot \sigma^{-1}
$$
(where the matrix $\sigma$ is given by (\ref{TensorAlgebra})).

In the special case of $N$ qubits (i.\,e.{} $n=2$) combining
Theorem~\ref{CatalanBasis} and Corollary~\ref{multiAlgebra} we obtain:

{\em Corollary \ref{coro8}.}
The set of matrices $T^{(N)}_{2,k}({\cal P}_k^N)$ is a vector space
basis of the algebra ${\cal A}_{2,k}^{(N)}$.

So far, we are able to compute a vector space basis for ${\cal
A}^{(N)}_{2,k}$ as follows:
\begin{itemize}
\item Generate the set ${\cal B}_k$ of all different binary trees.
\item Generate the set of permutations ${\cal P}_k$ obtained by
construction (\ref{tree2perm}).
\item For each $N$-tuple of permutations $(\pi_1,\ldots,\pi_N)$
apply the representation $T^{(N)}_{2,k}$, i.\,e., compute the tensor
product of the representations $T_{2,k}(\pi_\nu)$.
\end{itemize}
Instead of computing a matrix for each of the $(k!)^N$ tuples of
permutations in $(S_k)^N$, it is sufficient to consider only the
$C(k)^N$ permutations, implying a complexity reduction from
$O(k^{kN})$ to $O(4^{kN})$.

Using equation (\ref{trace_poly}), we get a set of polynomials
invariant under local transformations spanning the vector space
of homogeneous polynomial invariants of degree $k$.

For any $N$-tuple $\pi=(\pi_1,\ldots,\pi_N)\in(S_k)^N$ of
permutations we obtain a homogeneous invariant of degree $k$ given by
\begin{equation}\label{perminvariant}
f_{\pi_1,\ldots,\pi_N}(\rho_{i,j})
:=  \trace\bigl(T^{(N)}_{2,k}(\pi_1,\ldots,\pi_N)
        \cdot \rho^{\otimes k}\bigr).
\end{equation}

Clearly, there exist relations between the invariant polynomials
obtained from the tuple of permutations
$(\pi_1,\ldots,\pi_N)\in(S_k)^N$ (also with varying $k$). Some of
these relations can be expressed in terms of the permutations. This
allows a further reduction of the number of tuples of permutations
that have to be considered to compute all invariants.

{\em Theorem \ref{conjugation}.}
If $(\pi_1,\ldots,\pi_N)$ and $(\pi'_1,\ldots,\pi'_N)$ are
``simultaneously'' conjugated, i.\,e., there exists $\tau\in S_k$ such
that $\pi'_\nu=\tau^{-1}\pi_\nu\tau$ for all $\nu\in\{1,\ldots,N\}$,
then
$f_{\pi_1,\ldots,\pi_N}(\rho_{ij})=f_{\pi'_1,\ldots,\pi'_N}(\rho_{ij})$.

{\em Proof.} Simultaneous conjugation of the permutations $\pi_\nu$ by
$\tau$ corresponds to permuting the tensor factors in $\rho^{\otimes
k}$ by $\tau$, keeping $\rho^{\otimes k}$ fixed as a whole. Thus
$f_{\pi_1,\ldots,\pi_N}(\rho_{ij})$ does not change.

\qed
Next we give a condition on the permutations when an invariant can be
written as a product of invariants.

{\em Theorem \ref{transitive}.}
If the subgroup $H\le S_k$ generated by $\pi_1,\ldots,\pi_N$ is not
transitive, then the homogeneous invariant
$f_{\pi_1,\ldots,\pi_N}(\rho_{ij})$ of degree $k$ is a product of
invariants of smaller degree.

{\em Proof.} If $H$ is not transitive, it defines a non-trivial
partition of the set $\{1,\ldots,k\}$ into orbits. By simultaneous
conjugation using Theorem~\ref{conjugation}, we can assume that the
partition $\{1,\ldots,k_1\}$ and $\{k_1+1,\ldots,k\}$
respects the orbits. Thus, for $\nu=1,\ldots,N$ each permutation
$\pi_\nu$ can be written as a product $\pi_\nu=\pi'_\nu\cdot
\pi''_\nu$ with $\pi'_\nu\in S_{k_1}$ operating on $\{1,\ldots,k_1\}$
and $\pi''_\nu\in S_{k_2}$ operating on
$\{k_1+1,\ldots,k\}$ ($k_1+k_2=k$). Furthermore,
\begin{eqnarray*}
&&
f_{\pi_1,\ldots,\pi_N}(\rho_{ij})=
\trace\bigl(T^{(N)}_{2,k}(\pi_1,\ldots,\pi_N)
        \cdot \rho^{\otimes k}\bigr)\\
&&
\quad=
\trace\Bigl(\bigl(T^{(N)}_{2,k_1}(\pi'_1,\ldots,\pi'_N)
               \otimes T^{(N)}_{2,k_2}(\pi''_1,\ldots,\pi''_N)\bigr)
        \cdot \rho^{\otimes k}\Bigr)\\
&&\quad=
\trace\bigl(T^{(N)}_{2,k_1}(\pi'_1,\ldots,\pi'_N)
        \cdot \rho^{\otimes k_1}\bigr)\\
&&\qquad\qquad
  \cdot
  \trace\bigl(T^{(N)}_{2,k_2}(\pi''_1,\ldots,\pi''_N)
        \cdot \rho^{\otimes k_2}\bigr)\\
&&\quad=
f_{\pi'_1,\ldots,\pi'_N}(\rho_{ij})\cdot
f_{\pi''_1,\ldots,\pi''_N}(\rho_{ij}).
\end{eqnarray*}
\qed

For the case of two qubits, Table~\ref{numbers} shows the reduction of
the number of pairs of permutations to be considered using the
construction of permutations from binary trees,
Theorem~\ref{conjugation}, Theorem~\ref{transitive}, both Theorem
\ref{conjugation} and \ref{transitive}, and finally the combination of
Theorems \ref{conjugation}, \ref{transitive}, and \ref{melting}.
\begin{table}[hbt]
\small
\begin{tabular}{l|r|r|r|r|r|r}
$k$ & $(k!)^2$ & $C(k)^2$ & Th.~\ref{conjugation}
& Th.~\ref{transitive}
&\begin{tabular}{@{}c@{}}
        Th.~\ref{conjugation}\\
        Th.~\ref{transitive}
   \end{tabular}
&\begin{tabular}{@{}c@{}}
        Th.~\ref{conjugation}\\
        Th.~\ref{transitive}\\
        Th.~\ref{melting}
  \end{tabular}\\
\hline
1 & 1          & 1       & 1    & 1     & 1 & 1\\
2 & 4          & 4       & 4    & 3     & 3 & 2\\
3 & 36         & 25      & 10   & 15    & 6 & 3\\
4 & 576        & 196     & 36   & 97    & 20 & 10\\
5 & 14\,400      & 1\,764    & 114  & 733   & 60 & 22\\
6 & 518\,400     & 17\,424   & 496  & 6\,147  & 291 & 100\\
7 & 25\,401\,600   & 184\,041  & 2\,142 & 55\,541 & 1\,310 &361\\
8 &1\,625\,702\,400&2\,044\,900&10\,758&530\,773&6\,975&1\,717
\end{tabular}
\caption{Number of pairs $(\pi_1,\pi_2)\in(S_k)^2$ to be considered
for the construction of invariants using the different
theorems.\label{numbers}}
\end{table}

\section{Pure states and subspaces}\label{PureSubspace}
The technique to compute polynomials invariant under the action of
tensor products of unitary groups does not only apply to density
operators of mixed states, but also to subspaces and pure states. To
study non-local properties of subspaces with basis $\ket{\psi_i}$
(e.\,g.{} quantum error-correcting codes), one can use the invariants
of the corresponding projection operator
$P=\sum_i\ket{\psi_i}\bra{\psi_i}$. Pure states $\ket{\phi}$ can be
considered as one-dimensional subspace with projection operator
$P=\ket{\phi}\bra{\phi}$, or equivalently as a mixed state with
density operator $\rho=\ket{\phi}\bra{\phi}$.

In that situation, we have the additional relation $P^2=P$ (resp.{}
$\rho^2=\rho$) which can be used for a further reduction of the number
of permutations to be considered. The following theorem is quoted from
\cite{Rains97}, adding an explicit proof.

{\em Theorem \ref{melting}.}
Let $P$ be a projection operator. If for
$(\pi_1,\ldots,\pi_N)\in(S_k)^N$ there exist different numbers $l$ and
$m$ such that for each permutation $\pi_\nu$ we have $\pi_\nu(l)=m$,
then the invariant
$f_{\pi_1,\ldots,\pi_N}(P_{ij})=f_{\pi'_1,\ldots,\pi'_N}(P_{ij})$
where the permutations $\pi'_\nu\in S_{k-1}$ are obtained by
identifying the points $l$ and $m$ followed by a relabeling.

{\em Proof.} By Theorem~\ref{conjugation}, we can assume without loss
of generality that $l=1$ and $m=2$, i.\,e., $\pi_\nu(1)=2$ for all
permutations $\pi_\nu$. We will show that in the summation
(\ref{perminvariant}) there are entries of $P^2$ which can be replaced
by those of $P$. The entries of $P^{\otimes k}$ are of the form
$$
\left(P^{\otimes k}\right)_{
  (i^{(1)},\ldots,i^{(k)}),
  (j^{(1)},\ldots,j^{(k)})}
=P_{i^{(1)},j^{(1)}}\cdot\ldots\cdot P_{i^{(k)},j^{(k)}}.
$$
Here the indices $i^{(\mu)}$ are $N$-tuples
$(i_1^{(\mu)},\ldots,i_N^{(\mu)})$ (for a system with $N$
particles). The subscript of $i_{\nu}^{(\mu)}$ corresponds to the
$\nu$-th particle, whereas the superscript corresponds to the
$\mu$-th copy of the whole system. Left-multiplication by
$T_{n,k}^{(N)}(\pi_1,\ldots,\pi_2)$ permutes the rows of $P^{\otimes
k}$ yielding the matrix $M$ with entries
$$
M_{
  (i^{(1)},\ldots,i^{(k)}),
  (j^{(1)},\ldots,j^{(k)})}
=P_{i^{(\pi(1))},j^{(1)}}\cdot\ldots\cdot P_{i^{(\pi(k))},j^{(k)}}.
$$
The operation of $\pi=(\pi_1,\ldots,\pi_N)$ on the indices is given by
$i^{(\pi(\mu))}=(i_1^{(\pi_1(\mu))},\ldots,i_N^{(\pi_N(\mu))})$. Since
$\pi_\nu(1)=2$ we have
\begin{eqnarray*}
&&
M_{
  (i^{(1)},\ldots,i^{(k)}),
  (j^{(1)},\ldots,j^{(k)})}\\
&&\qquad
=P_{i^{(2)},j^{(1)}}\cdot
  P_{i^{(\pi(2))},j^{(2)}}\cdot\ldots\cdot P_{i^{(\pi(k))},j^{(k)}}.
\end{eqnarray*}
Now taking the trace results in
\begin{eqnarray}
&&
\sum_{j^{(1)},\ldots,j^{(k)}}
M_{
  (j^{(1)},\ldots,j^{(k)}),
  (j^{(1)},\ldots,j^{(k)})}\nonumber\\
&&\quad
=\sum_{j^{(1)},\ldots,j^{(k)}}
P_{j^{(2)},j^{(1)}}\cdot
  P_{j^{(\pi(2))},j^{(2)}}\cdot\ldots\cdot P_{j^{(\pi(k))},j^{(k)}}.
\label{traceprod}
\end{eqnarray}
Considering the summation over $j^{(2)}$ separately and using $P^2=P$,
we get
\begin{equation}\label{reduction}
\sum_{j^{(2)}} P_{j^{(\pi(2))},j^{(2)}}
P_{j^{(2)},j^{(1)}} =P_{j^{(\pi(2))},j^{(1)}}.
\end{equation}
Combining equations (\ref{traceprod}) and (\ref{reduction}) yields
\begin{eqnarray*}
&&\sum_{j^{(1)},\ldots,j^{(k)}}
M_{
  (j^{(1)},\ldots,i^{(k)}),
  (j^{(1)},\ldots,j^{(k)})}\\
&=&
\sum_{j^{(1)},j^{(3)},\ldots,j^{(k)}}
P_{j^{(\pi(2))},j^{(1)}}\cdot
  P_{j^{(\pi(3))},j^{(3)}}\cdot\ldots\cdot P_{j^{(\pi(k))},j^{(k)}}.
\end{eqnarray*}
Now the result follows immediately if we identify the points 1 and 2.

\qed

\section{The invariant ring of a two qubit system}\label{TwoQubit}
To illustrate the results, we consider the smallest non-trivial
example, a system of two qubits. Using our algorithm, we are able to
compute homogeneous invariants for each degree. As stated before, the
homogeneous invariants of fixed degree form a vector space. Therefore
it is sufficient to compute a basis for that vector space, e.\,g., a
maximal linearly independent set of homogeneous invariants. In order
to know how many invariants we need, we address the problem of
determining the dimension $d_k$ of the vector space of invariants of
degree $k$. Information about these dimensions is encoded in a formal
power series, the Molien series (cf.~\cite{Sturmfels})
$$
 P(z) := \sum_{k\geq 0} d_k z^k \in \Z[[z]].
$$
In case of a finitely generated algebra the Molien series turns out to
be a rational function (see e.\,g.~\cite{AtiyahMacdonald}). Thus it
can be expressed in a closed form with a finite number of terms. In
principle, the Molien series for the linear action
(\ref{lin_operation}) of a compact group $G$ on polynomials can be computed
by means of the following averaging formula:
\begin{equation}\label{molien}
 P(z) = \int_{g\in G}\frac{d\mu_G(g)}{{\det}(id - z g)}, 
\end{equation}
where $\mu_G$ is the normalized Haar measure of $G$.

In this paper, however, we are concerned with the action of
$G=SU(2)\times SU(2)$ on matrices by conjugation
(\ref{conj_operation}), for which the formula (\ref{molien}) does not
apply directly. Since the operation on $\rho$ given by
$$ 
\rho \mapsto (U_1 \otimes U_2)\cdot \rho\cdot (U_1\otimes U_2)^\dagger
\qquad\mbox{for $U_1, U_2 \in SU(2)$}
$$
is linear, we write $\rho$ as a vector $\vec{\rho}$ of length $n^2$
and obtain the representation
$$
\vec{\rho} \mapsto (U_1\otimes U_2 \otimes \overline{U_1} \otimes 
\overline{U_2}) \vec{\rho}
$$
where $\overline{U_i}$ denotes the complex conjugate of the matrix
$U_i$.

The integral (\ref{molien}) is simplified by means of the integral
formula of Weyl (cf.~\cite[$\S$26.2]{FultonHarris}) which allows to
perform an integration over the whole group in two steps. The first
step involves an integration over a maximal torus $T$ of the group $G$
and the second an integration on the residue classes $G/T$. Next, the
integral is transformed into a complex path integral which can be
solved by the theorem of residues. Finally, we end up with the Molien
series

{\footnotesize
\begin{eqnarray*}
P(z)&=&\frac{z^{10}-z^{8}-z^{7}+2z^{6}+2z^{5}+2z^{4}-z^{3}-z^{2}+1}{(z-1)^{10}(z+1)^{6}(z^2+1)^{2}(z^2+z+1)^{3}}\\
&=&1+z+4z^{2}+6z^{3}+16z^{4}+23z^{5}+52z^{6}+77z^{7}+150z^{8}\\
&&\phantom{1}
+224z^{9}+396z^{10}+583z^{11}+964z^{12}+1395z^{13}\\
&&\phantom{1}
+2180z^{14}+3100z^{15}+4639z^{16}+6466z^{17}+9344z^{18}\\
&&\phantom{1}
+12785z^{19}+17936z^{20}+24121z^{21}+33008z^{22}\\
&&\phantom{1}
+43674z^{23}+O(z^{24}).
\end{eqnarray*}
} 

The information about the dimensions of the vector spaces can now be
used to compute all invariants degree by degree. Having computed an
algebra basis for all invariants of degree less than $k$, homogeneous
invariants of degree $k$ are obtained by multiplying invariants of
smaller degrees that sum up to $k$. By computing the vector space
dimension of these invariants and comparing it to the dimension given
by the Molien series, we know how many linearly independent invariants
are missing. Next, these missing invariants are constructed from pairs
of permutations $\pi_\nu\in S_k$. It is sufficient to draw randomly
from the pairs of permutation remaining after
Theorem~\ref{conjugation} and Theorem~\ref{transitive} until the
vector space dimension given by the Molien series.

Using the computer algebra system {\sc Magma} \cite{Magma} we found 21
invariants corresponding to the permutations shown in
Table~\ref{permlist}. Furthermore, we were able to show that these
invariants generate all invariants up to degree 23. We conjecture that
any polynomial invariant of a two-qubit system can be expressed in
terms of these 21 invariants. It should be noted that there are
$43\,674$ linearly independent invariants of degree 23 and that the
invariants of degree 9 have more than $24\,000$ terms.

The Molien series provides also information about the maximal number
of algebraically independent invariants. This number is given by the
order of the pole $z=1$ of the Molien series \cite{Sturmfels}. For the
two-qubit system, there are 10 algebraically independent
invariants. Thus any 11 invariants fulfill a polynomial equation,
i.\,e., given numerical values for 10 algebraically independent
invariants, the values of the remaining invariants are some roots of
polynomials. But these values are not unique since none of the 21
invariants is a polynomial function of the others.

\begin{table}[hbt]
\begin{tabular}{c||c|c|r}
degree&$\pi_1$&$\pi_2$&\# terms\\
\hline
\hline
$1$&$id$&$id$&$4$\\
\hline
$2$&$id$&$(1\,2)$&$10$\\
$2$&$(1\,2)$&$id$&$10$\\
$2$&$(1\,2)$&$(1\,2)$&$10$\\
\hline
$3$&$(1\,2\,3)$&$(1\,2)$&$52$\\
$3$&$(1\,2\,3)$&$(1\,2\,3)$&$24$\\
\hline	
$4$&$(1\,2\,3\,4)$&$(1\,3)$&$110$\\
$4$&$(1\,2\,3\,4)$&$(1\,2\,3)$&$144$\\
$4$&$(1\,2\,3\,4)$&$(1\,2\,3\,4)$&$70$\\
$4$&$(1\,2\,3\,4)$&$(1\,2)(3\,4)$&$98$\\
\hline
$5$&$(1\,2\,3)(4\,5)$&$(1\,2\,3\,4\,5)$&$456$\\
\hline
$6$&$(1\,2\,3\,4\,5\,6)$&$(1\,2\,3\,5)$&$1\,334$\\
$6$&$(1\,2\,3)(4\,5)$&$(1\,2\,4\,5\,6)$&$1\,586$\\
$6$&$(1\,2\,3)(4\,5)$&$(1\,2\,3\,4\,5\,6)$&$1\,542$\\
$6$&$(1\,2\,3)(4\,5)$&$(1\,2\,3\,4)(5\,6)$&$1\,464$\\
\hline
$7$&$(1\,2\,3\,4)(5\,6\,7)$&$(1\,2\,4\,5\,6\,7)$&$4\,156$\\
$7$&$(1\,2\,3\,4)(5\,6\,7)$&$(1\,2\,6\,7)(3\,5)$&$4\,576$\\
\hline
$8$&$(1\,2\,3\,4\,5)(6\,7\,8)$&$(1\,2\,3\,5\,6\,7\,8)$&$10\,414$\\
$8$&$(1\,2\,3\,4\,5)(6\,7\,8)$&$(1\,2\,3\,7\,8)(4\,6)$&$11\,340$\\
\hline
$9$&$(1\,2\,3\,4\,5)(6\,7\,8)$&$(1\,2\,3\,6\,7\,8\,9)(4\,5)$&$24\,780$\\
$9$&$(1\,2\,3\,4\,5)(6\,7\,8\,9)$&$(1\,2\,3\,5\,6\,7\,8)$&$24\,168$
\end{tabular}
\caption{The invariants $f_{\pi_1,\pi_2}(\rho_{ij})$ corresponding to
these permutations (listed together with the degree and the number of
terms of the invariants) generate the polynomial invariants of a two
qubit system (at least) up to degree 23.\label{permlist}}
\end{table}

\section{Examples}\label{Examples}
\subsection{Characteristic polynomials}
As stated in the introduction, the coefficients of the characteristic
polynomial $\chi_\rho(X)$ of a density operator $\rho$ and of the
reduced density operators are invariant under local unitary
transformation. They can be expressed in terms of the invariants
$f_{\pi_1,\pi_2}$ presented in Table~\ref{permlist} as follows:

{\footnotesize
\begin{eqnarray*}
\chi_\rho(X)&=&
X^4
-f_{id,id}X^3
+\left(\frac{1}{2} f_{id,id}^2 - \frac{1}{2} f_{(1\,2),(1\,2)}\right)X^2\\
&&+\left(-\frac{1}{6} f_{id,id}^3 + \frac{1}{2} f_{id,id} f_{(1\,2),(1\,2)} - \frac{1}{3} f_{(1\,2\,3),(1\,2\,3)}\right)X\\
&&+\frac{1}{24} f_{id,id}^4 - \frac{1}{4} f_{id,id}^2 f_{(1\,2),(1\,2)} + 
    \frac{1}{3} f_{id,id} f_{(1\,2\,3),(1\,2\,3)}\\
&&\quad + \frac{1}{8} f_{(1\,2),(1\,2)}^2 - 
    \frac{1}{4} f_{(1\,2\,3\,4),(1\,2\,3\,4)}.
\end{eqnarray*}
}%
Here, the coefficient of $X^3$ in $\chi_\rho$ is a linear invariant
polynomial that equals the negative trace of $\rho$, and the constant
coefficient of $\chi_\rho$ is an invariant of degree four that equals
the determinant of $\rho$.

For the characteristic polynomials of the reduced density operators
$\rho^{(1)}:=\trace_2(\rho)$ and $\rho^{(2)}:=\trace_1(\rho)$ we
obtain
\begin{eqnarray*}
\chi_{\rho^{(1)}}(X)&=&X^2
-f_{id,id}X
+\frac{1}{2}\left(f_{id,id}^2-f_{(1\,2),id}\right)\quad\mbox{and}
\\
\chi_{\rho^{(2)}}(X)&=&X^2
-f_{id,id}X
+\frac{1}{2}\left(f_{id,id}^2 - f_{id,(1\,2)}\right).
\end{eqnarray*}

\subsection{Test for local equivalence}
Consider the following density operators:
{\small
$$
\rho_1=\frac{1}{800}\left(
\begin{array}{@{}cccc@{}}
        214 &   - 16-150\,i & 	  - 10 +8\,i& 	       3\\
 -16+150\,i & 	       226 & 	       13 &    10 -8\,i\\
  -10 -8\,i & 		13 & 	      234 &    16 + 150\,i\\
          3 & 	  10+8\,i &  16 -150\,i & 	     126
\end{array}
\right)
$$}%
and {\small
$$
\rho_2=\frac{1}{800}\left(
\begin{array}{@{}cccc@{}}
        214 & 	 - 16+150\,i & 	 10+8\,i & 	      -3\\
-16-150\,i & 	       226 & 	      -13 &    -10-8\,i\\
 10 -8\,i & 	       -13 & 	      234 &  16-150\,i\\
         -3 & 	  -10+8\,i &   16+150\,i & 	     126
\end{array}
\right)
$$}%
These density operators are (globally) conjugated to each other, i.\,e.,
$\rho_2=U \rho_1 U^\dagger$ where $U\in U(4)$.
%
Thus they have the same eigenvalues. Furthermore, the reduced density
operators are the same. Partial transposition of $\rho_1$ and $\rho_2$
yields positive operators $\rho_1^{T_i}$ and $\rho_2^{T_i}$ ($i=1,2$),
thus $\rho_1$ and $\rho_2$ are both separable \cite{separable}.
Again, the eigenvalues and the reduced density operators corresponding
to $\rho_1^{T_i}$ and $\rho_2^{T_i}$ are the same. Also, all
polynomial invariants up to degree five evaluate to the same number
for $\rho_1$ and $\rho_2$. Hence one might expect that $\rho_1$ and
$\rho_2$ are locally equivalent. 

But we have the following: The values of the invariants of degree six
corresponding to $\pi_1=(1\,2\,3)(4\,5)$, $\pi_2=(1\,2\,4\,5\,6)$ and
$\pi_1=(1\,2\,3)(4\,5)$, $\pi_2=(1\,2\,3\,4\,5\,6)$ differ showing
that $\rho_1$ and $\rho_2$ are not locally equivalent. Note that after
pre-computation of the invariants this can be decided just by
evaluating the invariants.

\section{Conclusion}
We have established a connection between local polynomial invariants
of quantum systems and permutations. The local invariants of an
$N$ particle system can be computed directly from $N$-tuples of
permutations. Theorems~\ref{conjugation} and \ref{transitive} allow a
reduction of the number of permutations to be considered. This makes
the computation of invariants feasible.

In the particular case of pure states and subspaces,
Theorem~\ref{melting} allows a further reduction of complexity. All
these theorems apply for subsystems of any dimension.

In the special case of qubits systems, we are able to construct a
vector space basis for the matrices directly in terms of permutations
derived from binary trees. It has to be investigated if there are
similar constructions for subsystems of dimension $n>2$.

Since our methods are not restricted to density operators, they can
also be used to study non-local properties of unitary
transformations. For example, it can be tested whether two quantum
circuits are equivalent with respect to conjugation by single qubit
gates.

\begin{acknowledgements}
The authors acknowledge fruitful discussions with G\"unter Mahler and
his co-workers at the University of Stuttgart and with J\"orn
M\"uller-Quade of IAKS, Karls\-ruhe. Furthermore, we would like to
thank Eric Rains for his comments. Part of this work was completed
during the 1997 Elsag-Bailey -- I.S.I. Foundation research meeting on
quantum computation. M.~R. is supported by ``Deutsche
Forschungsgemeinschaft, Graduiertenkolleg Beherrschbarkeit komplexer
Systeme, Universit\"at Karlsruhe'' under contract number DFG-GRK
209/3-98.
\end{acknowledgements}

\end{document}